\documentclass{article}

\usepackage{microtype}
\usepackage{graphicx}
\usepackage{subfigure,amssymb}
\usepackage{amsmath,tabularx}
\usepackage{booktabs} 

\usepackage{hyperref}

\usepackage[accepted]{icml2019}

\icmltitlerunning{MetricGAN: Generative Adversarial Networks based Black-box Metric Scores Optimization for Speech Enhancement}

\begin{document}

\twocolumn[
\icmltitle{MetricGAN: Generative Adversarial Networks based Black-box Metric Scores Optimization for Speech Enhancement}



\icmlsetsymbol{equal}{*}

\begin{icmlauthorlist}
\icmlauthor{Szu-Wei Fu}{ntu,as}
\icmlauthor{Chien-Feng Liao}{ntu,as}
\icmlauthor{Yu Tsao}{as}
\icmlauthor{Shou-De Lin}{ntu}
\end{icmlauthorlist}

\icmlaffiliation{ntu}{National Taiwan
University, Taiwan}
\icmlaffiliation{as}{Academia Sinica, Taiwan}

\icmlcorrespondingauthor{Yu Tsao}{yu.tsao@citi.sinica.edu.tw}
\icmlcorrespondingauthor{Shou-De Lin}{sdlin@csie.ntu.edu.tw}

\icmlkeywords{Machine Learning, ICML}

\vskip 0.3in
]

\renewcommand\arraystretch{1.3}


\printAffiliationsAndNotice{} 

\begin{abstract}
Adversarial loss in a conditional generative adversarial network (GAN) is not designed to directly optimize evaluation metrics of a target task, and thus, may not always guide the generator in a GAN to generate data with improved metric scores. To overcome this issue, we propose a novel MetricGAN approach with an aim to optimize the generator with respect to one or multiple evaluation metrics. Moreover, based on MetricGAN, the metric scores of the generated data can also be arbitrarily specified by users. We tested the proposed MetricGAN on a speech enhancement task, which is particularly suitable to verify the proposed approach because there are multiple metrics measuring different aspects of speech signals. Moreover, these metrics are generally complex and could not be fully optimized by $L_{p}$ or conventional adversarial losses.
\end{abstract}

\section{Introduction}
\label{submission}

Generative adversarial networks (GANs) \cite{goodfellow2014generative} has shown its powerful generative ability in many different applications. In particular, for conditional GANs (CGANs)  \cite{mirza2014conditional}, in addition to the adversarial loss, there is an $L_{p}$ loss, to guide the learning of generators. Ideally, the adversarial loss should make generated data indistinguishable from real (target) data. However, some applications of image \cite{ledig2017photo,wang2018esrgan} and speech processing \cite{pandey2018adversarial,wang2018supervised,donahue2018exploring, michelsanti2017conditional} show that this loss term provides very marginal improvement (sometimes even degrade the performance) in terms of objective evaluation scores (in the case of image processing, the subjective score can be improved). For instance, Donahue et al. \yrcite{donahue2018exploring} applied CGAN on speech enhancement (SE) for automatic speech recognition (ASR); however, the following conclusion was obtained: "Our experiments indicate that, 
for ASR, simpler regression approaches may be preferable
to GAN-based enhancement." This may be because the method that the discriminator uses to judge whether each sample is real or fake is not fully related to the metrics that we consider. In other words, similar to  $L_{p}$ loss, the way the adversarial loss guides the generator to generate data is still not matched to the evaluation metrics. We call this problem discriminator-evaluation mismatch (DEM). 
In this study, we propose a novel MetricGAN to solve this problem. We tested the proposed approach on the SE task because the metrics for SE are generally complex and difficult to directly optimize or adjust.

For human perception, the primary goal of SE is to improve the intelligibility and quality of noisy speech \cite{benesty2015speech}. To evaluate a SE model in different aspects, several objective metrics have been proposed. Among the human perception-related objective metrics, the perceptual evaluation of speech quality (PESQ) \cite{rix2001perceptual} and short-time objective intelligibility (STOI) \cite{taal2011algorithm} are two popular functions to evaluate speech quality and intelligibility, respectively. The design of these two metrics considers human auditory perception and has shown higher correlation to subjective listening tests than simple $L_{1}$ or $L_{2}$ distance between clean and degraded speech. 

In recent years, various deep learning-based models have been developed for SE \cite{lu2013speech,xu2014experimental,wang2014training,xu2015regression, ochiai2017multichannel,luo2018tasnet,grais2018raw,germain2018speech,chai2018error, choi2018phase}. 
Most of these models were trained in a supervised fashion by preparing pairs of noisy and clean speeches. The deep models were then optimized by minimizing the distance between generated speech and clean speech. However, the distance (objective function) is usually based on simple $L_{p}$ loss (where $p$ = 1 or 2), which does not reflect human auditory perception or ASR  accuracy  \cite{bagchi2018spectral} well. In fact, several researches have indicated that an enhanced speech with a smaller $L_{p}$ distance, does not guarantee a higher quality or intelligibility score \cite{fu2018end,koizumi2018dnn}.


Therefore, optimizing the evaluation metrics (i.e., STOI, PESQ, etc.) may be a reasonable direction to connect the model training with the goal of SE. Some latest studies \cite{fu2018end,koizumi2018dnn,zhang2018training,zhao2018perceptually,naithani2018deep, kolbaek2018monaural, venkataramani2018performance,venkataramani2018end,zhao2018convolutional} have focused on STOI score optimization to improve speech intelligibility. A waveform based utterance-level enhancement manner is proposed to optimize the STOI score \cite{fu2018end}. The results of a listening test showed that by combining STOI with MSE as an objective function, the speech intelligibility can be further increased. 
On the other hand, because the PESQ function is not fully differentiable and significantly more complex compared with STOI, only few \cite{koizumi2018dnn,zhang2018training,koizumi2017dnn, martin2018deep} have considered it as an objective function. Reinforcement learning (RL) techniques such as deep Q-network (DQN) and policy gradient were employed to solve non-differentiable  problems, as \cite{koizumi2017dnn} and \cite{koizumi2018dnn}, respectively. 


In summary, the abovementioned existing techniques can be categorized into two types depending on whether the details of evaluation metrics have to be obtained: (1) white-box: these methods approximate the complex evaluation metrics with a hand-crafted, simpler one; thus, it is differentiable and easy to be applied as a loss function. However, the details of the metrics have to be known; (2) black-box: these methods mainly treat the metrics as a reward and apply RL-based techniques to increase the scores. However, because of less efficiency in training, most of them have to be pre-trained by conventional supervised learning.
 
 In this study, to solve the drawbacks of the abovementioned methods and the DEM problem, the discriminator in GAN is associated with the evaluation metrics of interest (Although these evaluation functions are complex, Fu et al. \yrcite{fu2018quality} showed that they can be approximated by neural networks). In particular, when training the discriminator, instead of always giving a false label (e.g., "0") to the generated speech, the labels of  MetricGAN are given according to the evaluation metrics. Therefore, the target space of discriminator transforms from discrete (1 (true) or 0 (false)) to continuous (evaluation scores). Through this modification, the discriminator can be treated as a learned surrogate of the evaluation metrics. In other words, the discriminator iteratively estimates a surrogate loss that approximates the sophisticated metric surface, and the generator uses this surrogate to decide a gradient
direction for optimization. Compared with previous existing methods, the main advantages of MetricGAN are as follows:
 
(1) The surrogate function (discriminator) of the complex evaluation metrics is learned from data. In other words, it is still in a black-box setting and no computational details of the metric function have to be known.

(2) Experiment result shows that the training efficiency of MetricGAN to increase metric score is even higher than conventional supervised learning with $L_{p}$ loss.

(3) Because the label space of the discriminator is now continuous, any desired metric scores can be assigned to the generator. Therefore, MetricGAN has the flexibility to generate speech with specific evaluation scores. 

(4) Under some non-extreme conditions, MetricGAN can even achieve multi-metrics assignments by employing multiple discriminators.


\section{CGAN for SE}
GAN has recently attracted a significant amount of attention in the community. By employing an alternative mini-max training scheme between a generator network ($G$) and a discriminator network ($D$), adversarial training can model the distribution of real data. One of its applications is to serve as a trainable objective function for a regression task. Instead of explicitly minimizing the $L_{p}$ losses, which may cause over smoothing problems, $D$ provides a high-level abstract measurement of ”realness” \cite{liao2018noise}. 

In the applications of GAN on SE, CGAN is usually employed to generate enhanced speech. To achieve this, $G$ is trained to map noisy speech $x$ to its corresponding clean speech $y$ by minimizing the following loss function (as in \cite{pascual2017segan}. The least-squares GAN (LSGAN) approach \cite{mao2017least} is used with binary coding (1 for real, 0 for fake)):
\begin{equation}
L_{G (CGAN)}= \mathbb{E}_{x} [\lambda(D(G(x), x) - 1)^2]+ ||G(x) - y||_{1}
\end{equation}

Because $G$ usually simply learned to ignore the noise prior $z$ in the CGAN \cite{isola2017image}, we directly neglected it here. The first term in Eq. (1) is called adversarial loss for cheating $D$ with a weighting factor $\lambda$. The goal of $D$ is to distinguish between real data and generated data by minimizing the following loss function:
\begin{equation}
L_{D (CGAN)}= \mathbb{E}_{x,y} [(D(y, x) - 1)^2+(D(G(x), x) -0)^2]
\end{equation}

We argue that to optimize the metric scores, the training of $D$ should be associated with the metric.  

\section{MetricGAN}
\subsection{Associating the Discriminator with the Metrics}
The main difference between the proposed MetricGAN and the conventional CGAN is how the discriminator is trained. Here, we first introduce a function $Q (I)$ to represent the evaluation metric to be optimized, where $I$ is the input of the metric. For example, for PESQ and STOI, $I$ is the pair of speech that we want to evaluate and the corresponding clean speech $y$. Therefore, to ensure that $D$ behaves similar to $Q$, we simply modify the objective function of $D$:
\begin{equation}
\begin{aligned}
L_{D (MetricGAN)}&= \mathbb{E}_{x,y} [(D(y, y) - Q(y, y))^2
\\ &+(D(G(x), y) - Q(G(x), y))^2]
\end{aligned}
\end{equation}

Because we can always map $Q$ to $Q'$, which is between 0 and 1 (here, 1 represents the best evaluation score), Eq. (3) can be reformulated as
\begin{equation}
\begin{aligned}
L_{D (MetricGAN)}&= \mathbb{E}_{x,y} [(D(y, y) - 1)^2
\\ &+(D(G(x), y) - Q'(G(x), y))^2]
\end{aligned}
\end{equation}

where 0 $\leq$ $Q'(G(x), y)$ $\leq$ 1. There are two main differences between Eq. (4) and Eq. (2): 

1.) In CGAN, as long as the data is generated, its label for $D$ is always a constant 0. However, the target label of the generated data in our MetricGAN is based on its metric score. Therefore, $D$ can evaluate the degree of realness (clean speech), instead of just distinguishing real and fake. (Therefore, maybe "$D$" should be called an evaluator; however, here we just follow the convention of GAN.)

2.) The condition used in the $D$ of CGAN is the noisy speech $x$, which is different from the condition used in the proposed MetricGAN (clean speech $y$). This is because we want $D$ and $Q$ to have similar behavior. Therefore, the input argument of $D$ is chosen to be the same as $Q$.

\subsection{Continuous Space of the Discriminator Label}
The training of $G$ is similar to Eq. (1). However, we found that the gradient provided by $D$ in our MetricGAN is more efficient than the $L_{p}$ loss. Therefore, the training of $G$ can completely rely on the adversarial loss :
\begin{equation}
L_{G (MetricGAN)}= \mathbb{E}_{x} [(D(G(x), y) - s)^2]
\end{equation}

where $s$ is the desired assigned score. For example, to generate clean speech, we can simply assign $s$ to be 1. On the contrary, we can also generate more noisy speech by assigning a smaller $s$. This flexibility is caused by the label of the generated speech in $D$, which is now continuous and related to the metric. Unlike surrogate loss learning in the multi-class classification \cite{hsieh2018deep}, because the output space of our $G$ is continuous, the local neighbors need not be explicitly selected to learn the behavior of metric surface.

\begin{figure}[ht]
\vskip 0.1in
\begin{center}
\centerline{\includegraphics[width=\columnwidth]{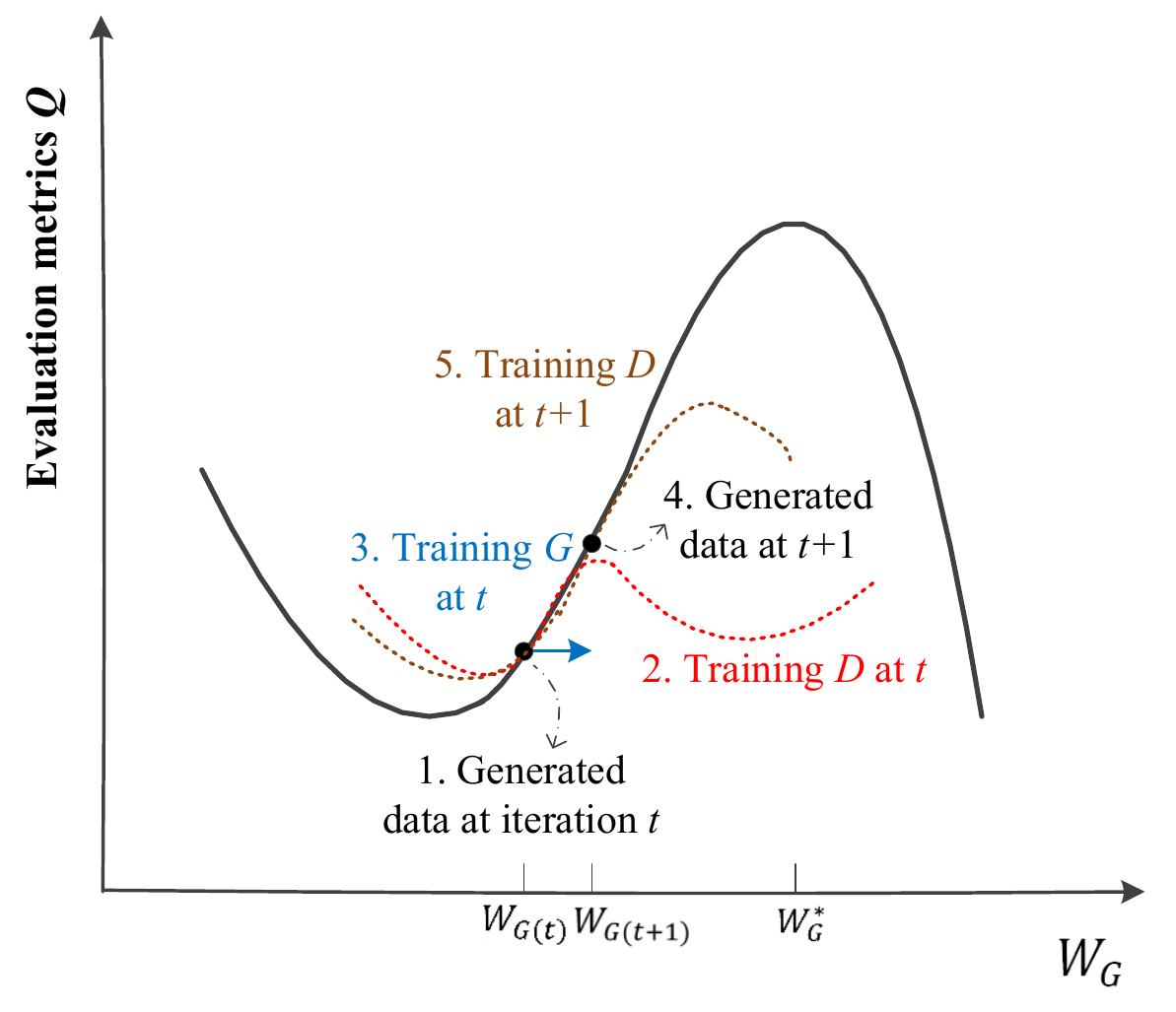}}
\vskip -0.1in
\caption{Learning process of MetricGAN to optimize the evaluation metric $Q$ (the horizontal axis represents the weights of $G$). In each iteration, there are three steps. First, some data are generated by $G$ with weights $W_{G(t)}$. Second, training $D$ to have similar behavior as metric $Q$ in those points. Third, training $G$ according to the gradient provided by $D$ (cheat $D$).}
\label{fig:MetricGAN_explain}
\end{center}
\vskip -0.2in
\end{figure}

\subsection{Explanation of MetricGAN}
In MetricGAN, the target of G is to cheat D to reach specified score, and D tries to not be cheated by learning the true score. Here, we also explain the learning process of MetricGAN in a different manner. As shown in Figure~\ref{fig:MetricGAN_explain}, training of $D$ can be treated as learning a local surrogate of $Q$; and training of $G$ is to adjust its weights $W_G$ toward the optimum value of $D$. Because $D$ may only approximate $Q$ well in the observed region \cite{fu2019learning}, this learning framework should be alternatively trained until convergence. 

\section{Experiments}
\subsection{Network Architecture}
The input features $x$ for $G$ is the normalized noisy magnitude spectrogram utterance. The generator used in this experiment is a BLSTM \cite{weninger2015speech} with two bidirectional LSTM layers, each with 200 nodes, followed by two fully connected layers, each with 300 LeakyReLU nodes and 257 sigmoid nodes for mask estimation, respectively. When this mask (between 0 to 1) is multiplied with the noisy magnitude spectrogram, the noise components should be removed. In addition, as reported in \cite{koizumi2018dnn}, to prevent musical noise, flooring was applied to the estimated mask before T-F-mask processing. Here, we used the lower threshold of the T-F mask as 0.05.

\begin{table*}[!t]
\caption{Performance comparisons of different loss functions in terms of PESQ and STOI (* represents pre-trained from another model). }
\vskip -0.1in
\label{tab:different Loss Results}
\begin{center}
\begin{small}
\begin{tabular}{c||c|c||c|c||c|c||c|c||c|c||c|c}
\toprule
&\multicolumn{2}{c||}{\textbf{Noisy}} & \multicolumn{2}{c||}{\textbf{IRM (L1)}} & \multicolumn{2}{c||}{\textbf{IRM (CGAN)}} & \multicolumn{2}{c||}{\textbf{PE policy grad*(P)}} & \multicolumn{2}{c||}{\textbf{MetricGAN (P)}} & \multicolumn{2}{c}{\textbf{MetricGAN (S)}}\\
\hline
SNR (dB) & PESQ & STOI & PESQ & STOI & PESQ & STOI & PESQ & STOI & PESQ & STOI & PESQ & STOI \\
\hline
12	&2.375	&0.919	&2.913	&0.935	&2.879	&0.936	&\textbf{2.995}
&0.927 &2.967	&0.936	&2.864	&\textbf{0.939}
\\

6	&1.963	&0.831	&2.52	&0.878	&2.479	&0.876	&2.595
&0.869 &\textbf{2.616}	&0.881	&2.486	&\textbf{0.885} \\

0	&1.589	&0.709	&2.086	&0.787	&2.053	&0.786	&2.144
&0.776 &\textbf{2.200}	&0.796	&2.086	&\textbf{0.802} \\

-6	&1.242	&0.576	&1.583	&0.655	&1.551	&0.653	&1.634
&0.644 &\textbf{1.711}	&0.668	&1.599	&\textbf{0.679}\\

-12	&0.971	&0.473	&1.061	&0.508	&1.046	&0.507	&1.124
&0.500 &\textbf{1.169}	&0.521	&1.090	&\textbf{0.533}
\\
\hline
Avg.	&1.628	&0.702	&2.033	&0.753	&2.002	&0.751	&2.098
&0.743 &\textbf{2.133}	&0.760	&2.025	&\textbf{0.768}\\
\bottomrule
\end{tabular}
\end{small}
\end{center}
\vskip -0.1in
\end{table*}

The discriminator herein is a CNN with four two-dimensional (2-D) convolutional layers with the number of filters and kernel size as follows: [15, (5, 5)], [25, (7, 7)], [40, (9, 9)], and [50, (11, 11)]. To handle the variable-length input (different speech utterance has different length), a 2-D global average pooling layer was added such that the features can be fixed at 50 dimensions (50 is the number of feature maps in the previous layer). Three fully connected layers were added subsequently, each with 50 and 10 LeakyReLU nodes, and 1 linear node. In addition, to make $D$ a smooth function (we do not want a small change in the input spectrogram can result in a significant difference to the estimated score), it is constrained to be 1-Lipschitz continuous by spectral normalization \cite{miyato2018spectral}.  Our preliminary experiments found that adding this constraint can stabilize the training of $D$. All models are trained using Adam \cite{kingma2014adam} with ${\beta_1}$ = 0.9  and ${\beta_2}$ = 0.999.


\subsection{Experiment on the TIMIT Dataset}
In this section, we show the experiments about PESQ and STOI scores.  PESQ was designed to evaluate the quality of processed speech, and the score ranges from -0.5 to 4.5. STOI was designed to compute the speech intelligibility, and the score ranges from 0 to 1. Both the two metrics are the higher the better. 

\begin{figure}[ht]
\vskip -0.1in
\begin{center}
\centerline{\includegraphics[width=\columnwidth]{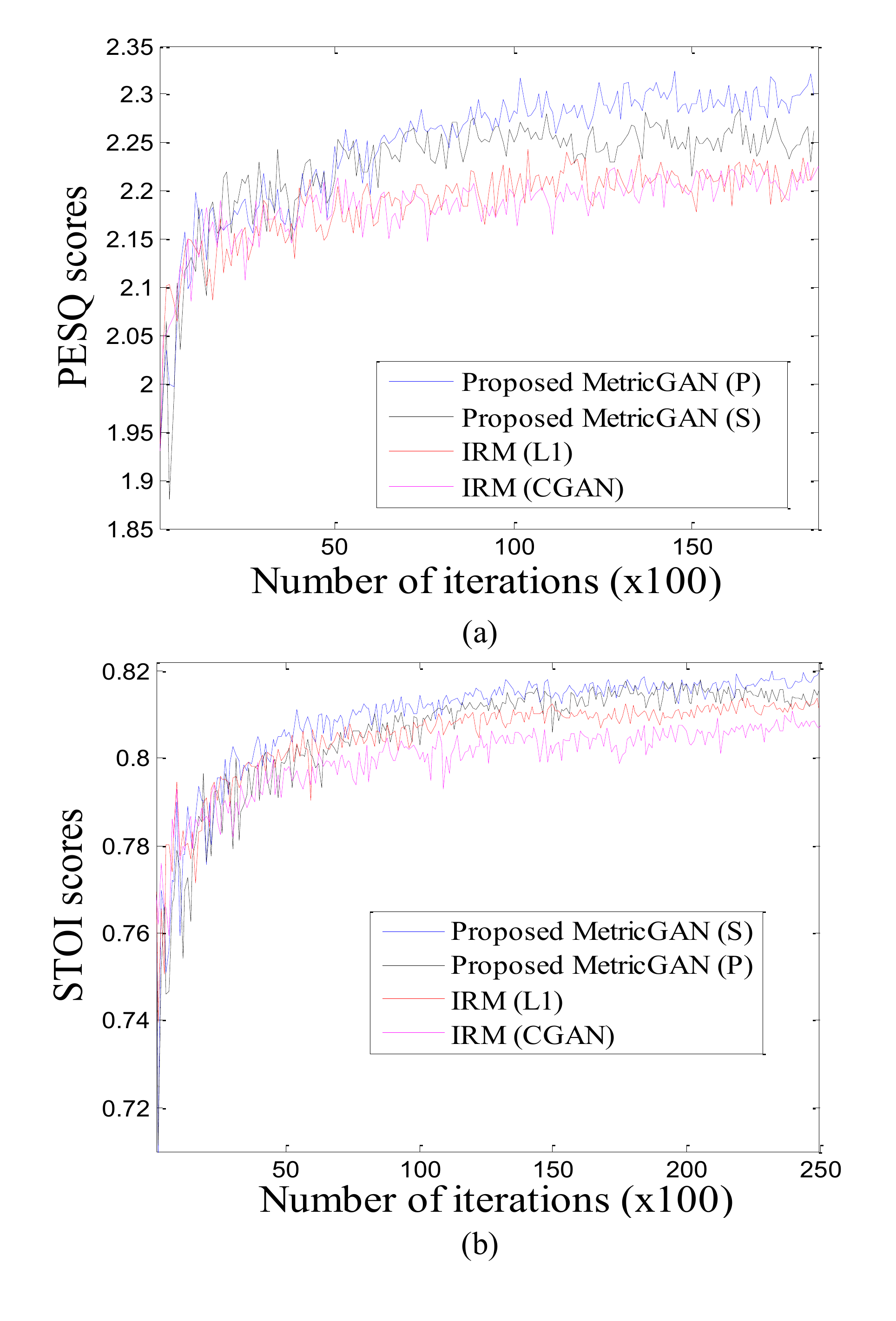}}
\vskip -0.2in
\caption{Learning curve of different objective functions evaluated on the validation set (structure of $G$ is fixed). In terms of: (a) PESQ score and (b) STOI score.}
\label{fig:learning_process}
\end{center}
\vskip -0.2in
\end{figure}

\begin{figure}[ht]
\vskip 0.05in
\begin{center}
\centerline{\includegraphics[width=\columnwidth]{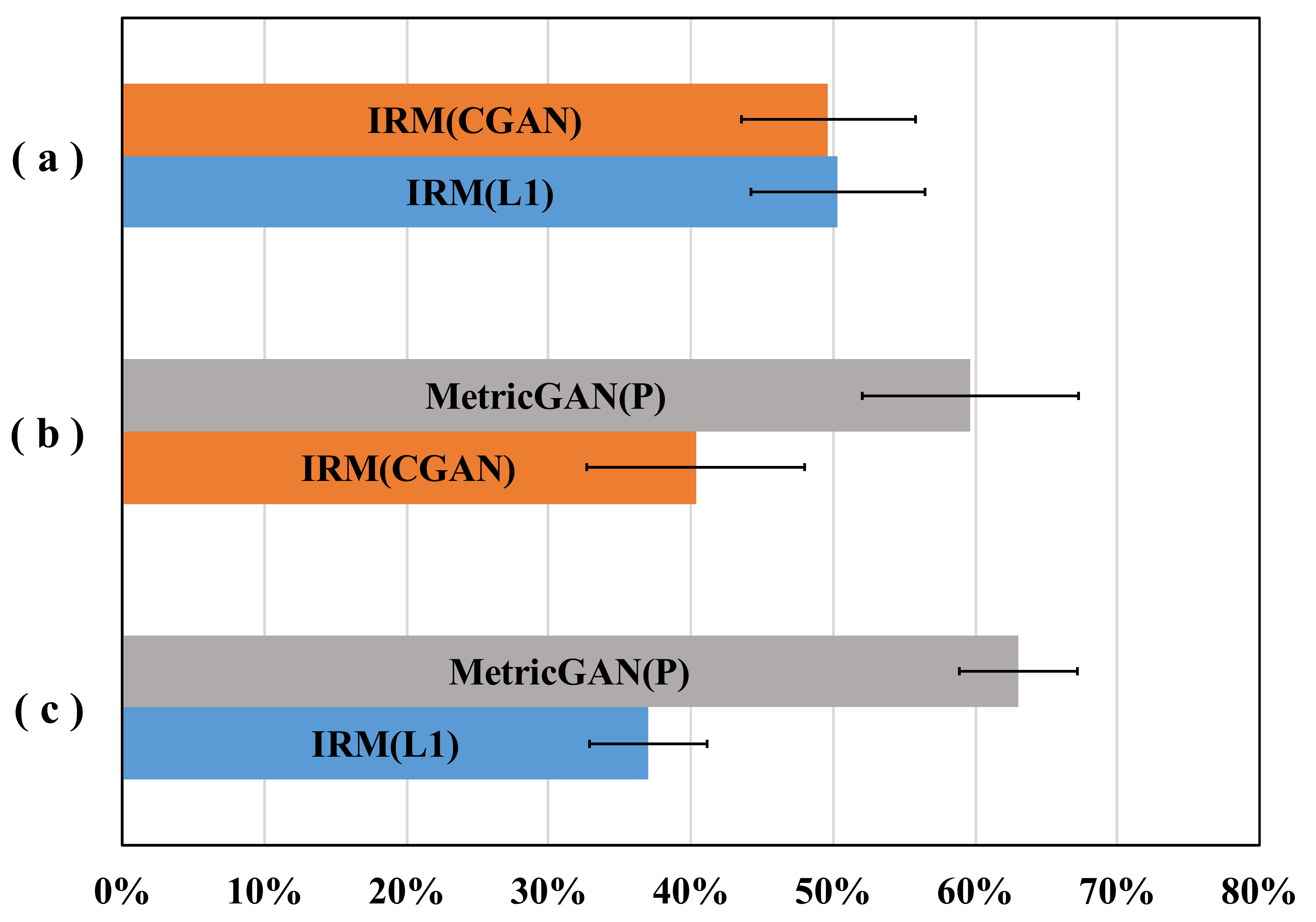}}
\vskip 0.0in
\caption{Results of AB preference test (with 95\% confidence intervals) on speech quality compared between proposed \textbf{MetricGAN(P)} and the two baseline models.}
\label{fig:abtest}
\end{center}
\vskip -0.2in
\end{figure}

\begin{figure*}[ht]
\begin{center}
\centerline{\includegraphics[width=17cm]{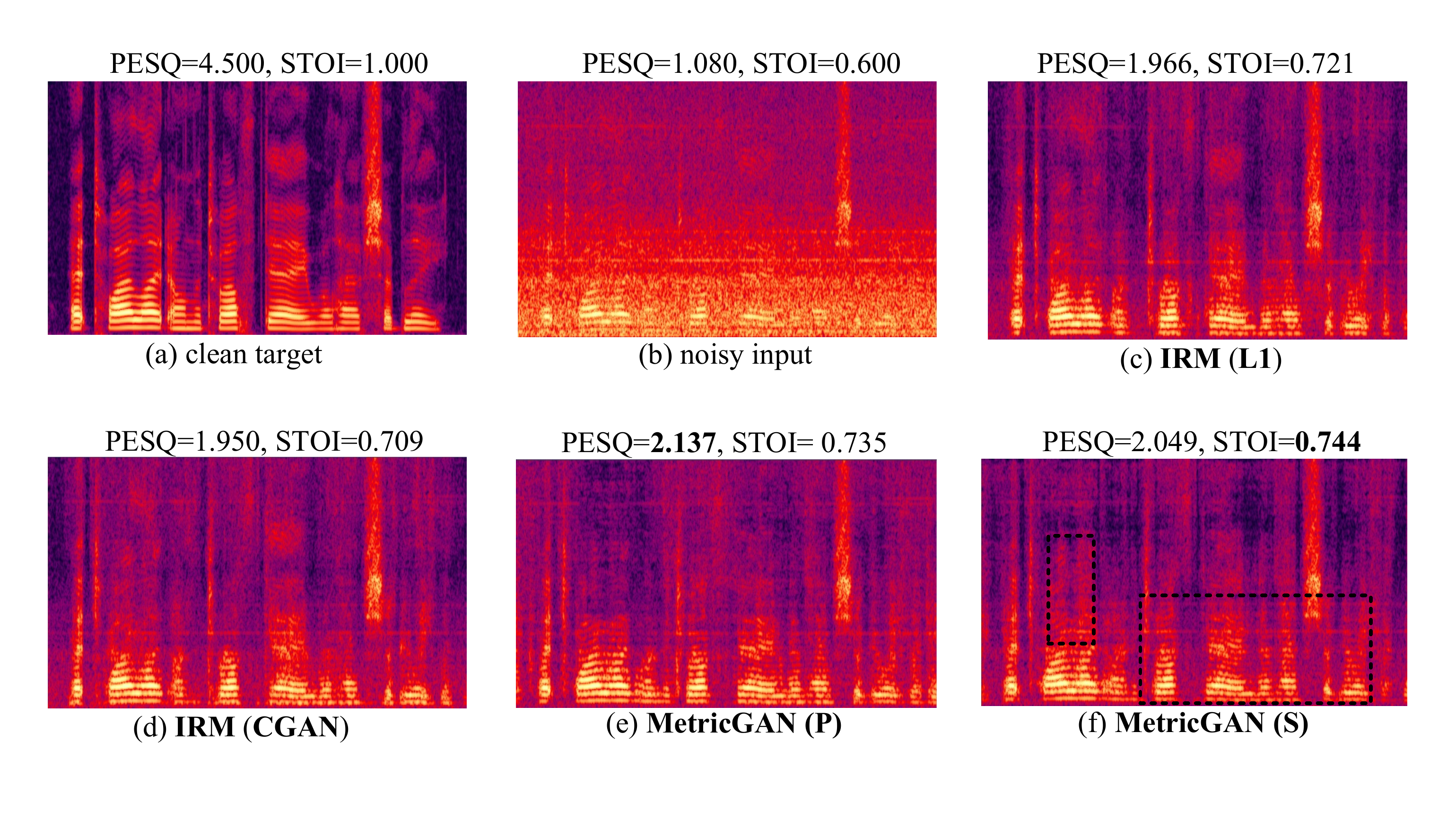}}
\vskip -0.3in
\caption{Spectrograms of a TIMIT utterance in the teset set: (a) clean target, (b) noisy speech (engine noise at 0 dB). (c) to (f): enhanced speech with different loss functions. }
\label{fig:demo_different_loss}
\end{center}
\vskip -0.2in
\end{figure*}

\subsubsection{Dataset}
In this experiments, the TIMIT corpus \cite{garofolo1988getting} was used to prepare the training, validation, and test sets. 300 utterances were randomly selected from the training set of the TIMIT database for training in this experiment. These utterances were further corrupted with 10 noise types (crowd, 2 machine, alarm and siren, traffic and car, animal sound, water sound, wind, bell, and laugh noise) from \cite{hu100}, at five SNR levels (from -8 dB to 8 dB with steps of 4 dB) to form 15000 training utterances. To monitor the training process and choose the hyperparameters, we randomly selected another clean 100 utterances from the TIMIT training set to form our validation set. Each utterance was further corrupted with one of the noise types (different from those already used in the training set) from \cite{hu100} at five different SNR levels (from -10 dB to 10 dB with steps of 5 dB). To evaluate the performance of different training methods, 100 utterances from the TIMIT test set were randomly selected as our test set. These utterances were mixed with four unseen noise types (engine, white, street, and baby cry), at five SNR levels (-12 dB, -6 dB, 0 dB, 6 dB, and 12 dB). In summary, 2000 utterances exist in the test set.

\subsubsection{Objective Evaluation with Different Loss Functions}
In this experiment, to evaluate the performance of different objective functions, the structure of $G$ is fixed and trained with different losses. As one of our baseline models, we adopt ideal ratio mask (IRM) \cite{narayanan2013ideal} based mask estimation with $L_{1}$ loss (denoted as \textbf{IRM (L1)}). The other baseline (denoted as \textbf{IRM (CGAN)}) is the CGAN with the loss function of $G$ shown in Eq. (1). Compared to \textbf{IRM (L1)}, \textbf{IRM (CGAN)} has an additional adversarial loss term with $\lambda$ = 0.01 as in \cite{bagchi2018spectral,pascual2017segan}. A parameter exploring policy gradients \cite{sehnke2010parameter} based black-box optimization, which is similar to the one used in \cite{zhang2018training}, is also compared. However, we found that this method is very sensitive to the hyperparameters (e.g., weight initialization, step size of jitter, etc.). We could only obtain improved results for PESQ optimization (denoted as \textbf{PE policy grad (P)}). In addition, because of the lower training efficiency, its generator was first pre-trained from \textbf{IRM (L1)}. The proposed MetricGAN with PESQ or STOI metric as $Q$, is indicated as \textbf{MetricGAN (P)} and \textbf{MetricGAN (S)}, respectively.

Table~\ref{tab:different Loss Results} presents the results of the average PESQ and STOI scores on the test set for the baselines and proposed methods. From this table, we can first observe that the performance of \textbf{IRM (CGAN)} is similar to or slightly worse than the simple \textbf{IRM (L1)}, which is in agreement with the results presented in previous papers.
\cite{pandey2018adversarial,donahue2018exploring}. This implies that the adversarial loss term used to cheat $D$ is not helpful in this application. One possible reason for this result may be that the decision boundary of $D$ is very different from the metrics we consider. We also attempted to train \textbf{IRM (CGAN)} with larger $\lambda$; however, their evaluation scores were worse than the reported scores. Although \textbf{PE policy grad (P)} can obtain some PESQ scores improvements, the STOI scores decreased compared to its initialization, \textbf{IRM (L1)}. On the contrary, when we employed PESQ as $Q$ in our MetricGAN, it could achieve the highest PESQ scores among all the models with the second highest STOI score. Note that unlike $L_p$ loss, the loss function of $G$ in MetricGAN is Eq. (5), and there is no specific target for each T-F bin. In terms of the STOI score, \textbf{MetricGAN (S)} outperforms the other models, and the improvement is most evident for the low SNR conditions (where speech intelligibility improvement is most critical).

In addition to the final results of the test set, the learning process of different loss functions evaluated on the validation set are also presented in Figure~\ref{fig:learning_process}. For both the scores, we can observe that the learning efficiency (in terms of the number of iterations) of MetricGAN is higher than the others. This implies that the gradient provided by $D$ (surrogate of $Q$) is the most accurate toward the maximum value of $Q$. However, if the $Q$ used to train MetricGAN does not match the evaluation metric, the performance is sub-optimal. Therefore, the information from $Q$ is important; our preliminary experiment also shows that without $Q$, the learning cannot converge. The conventional adversarial loss term in \textbf{IRM (CGAN)} is not helpful for improving the scores and training efficiency.

Finally, an example of the enhanced spectrograms by different training objective functions are shown in Figure~\ref{fig:demo_different_loss}. The spectrogram generated by \textbf{IRM (CGAN)} is similar to that of \textbf{IRM (L1)}. If we simply increase the weight $\lambda$ of the adversarial loss term in Eq.(1), some unpleasant artifacts begin to appear (this is not shown here, owing to limited space). Interestingly, in comparison to others, the spectrogram (f) generated by \textbf{MetricGAN (S)} can best recover the speech components with clear structures (as shown by the black-dashed rectangles) and hence, obtain the highest STOI score.

\subsubsection{Subjective Evaluation}
To evaluate the perceptual quality of the enhanced speech, we conducted AB preference tests to compare the proposed method with the baseline models. Three pairs of listening tests were conducted: \textbf{IRM(CGAN)} versus \textbf{IRM(L1)}, \textbf{MetricGAN(P)} versus \textbf{IRM(CGAN)}, and \textbf{MetricGAN(P)} versus \textbf{IRM(L1)}. Each pair of samples are presented in a randomized order. For each listening test, 20 sample pairs were randomly selected from the test set; 15 listeners participated. Listeners were instructed to select the sample with the better quality. The stimuli were played to the subjects in a quiet environment through a set of Sennheiser HD headphones at a comfortable listening level. In Figure \ref{fig:abtest} (a), we can observe that the preference score between \textbf{IRM (L1)} and \textbf{IRM (CGAN)} overlap in the confidence interval, which is in agreement with the result of the objective evaluation. Further, as shown in Figure \ref{fig:abtest} (b) and Figure \ref{fig:abtest} (c), \textbf{MetricGAN(P)} significantly outperforms both baseline systems, without an overlap in the confidence intervals.

\subsubsection{Assigning Any Desired Score to the Generator}
\begin{figure*}[ht]
\begin{center}
\centerline{\includegraphics[width=18cm]{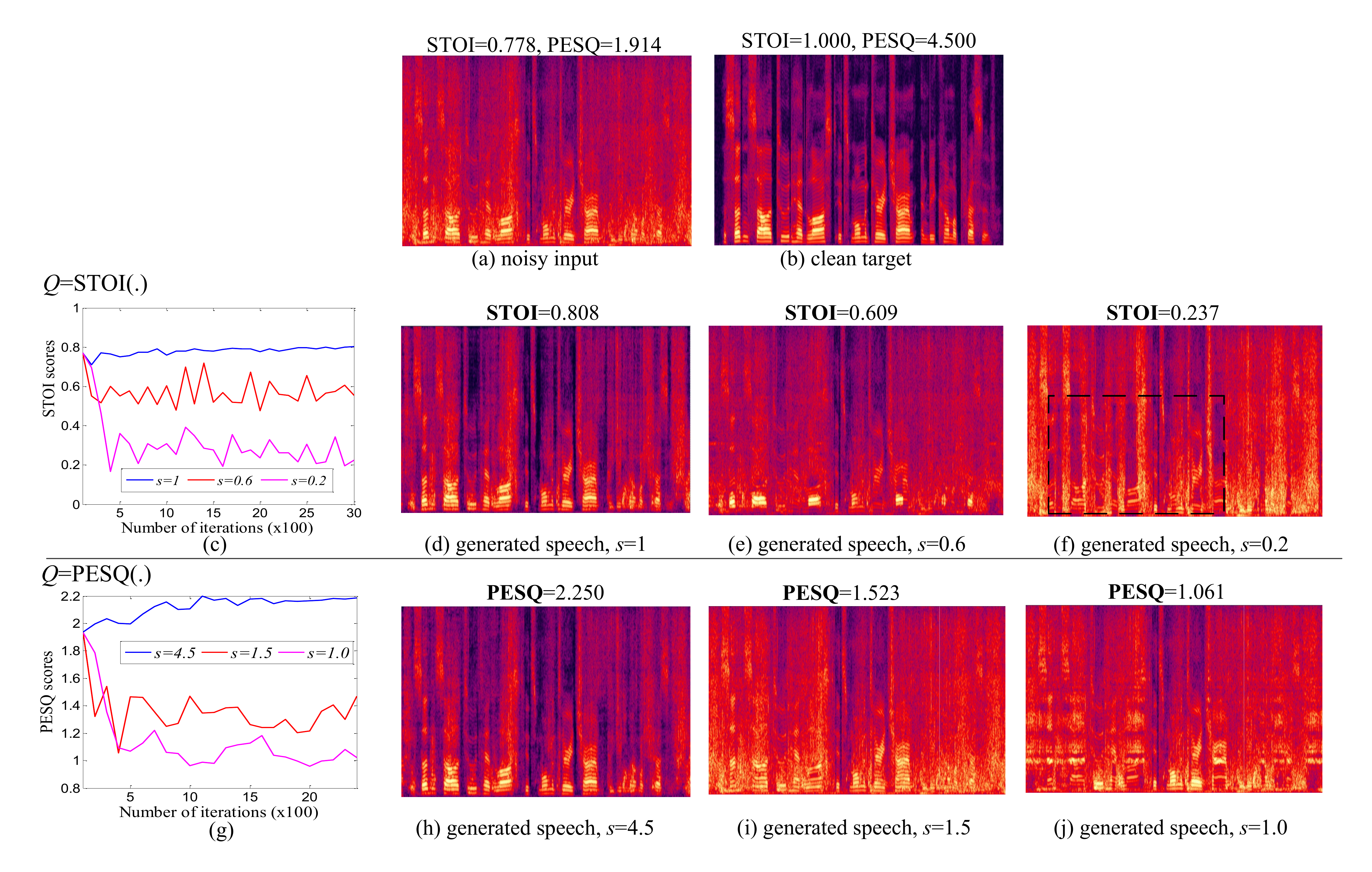}}
\vskip -0.2in
\caption{Results of assigning different $s$ to Eq. (5) for the generator training. Note that the learning curves of generating clean speech in (c) and (g) are not yet converged. For more complete learning processes, please refer to Figure~\ref{fig:learning_process}.}
\label{fig:demo_different_score}
\end{center}
\vskip -0.2in
\end{figure*}

Because the label of $D$ in the conventional GAN is in a discrete space, there is no guarantee that the generated data can gradually improve toward real data when the label assigned to $G$ (i.e., the constant in the first term of Eq.(1)) increases  from 0 (fake) toward 1 (real). For example, the generated data from label 0.9 is not necessarily better (more like real data) than that from label 0.8.
However, as pointed out in section 3.2, because the output of $D$ in MetricGAN is continuous according to $Q$, we can assign any desired score during the training of $G$ as in Eq. (5). Therefore, different $s$ in Eq. (5) correspond to generated speech with different qualities. Interestingly, setting $s$ as a small value can convert the generator from a speech enhancement model to a speech degradation model. This provides us with another method to understand the factors that affect the metric. To achieve this, a uniform mask constraint (penalize estimated mask away from 0.5) was also applied to $G$ so that $G$ has to choose the most efficient way to attain the assigned score $s$ without significantly changing the initialized mask. (Owing to the sigmoid activation used in the output layer of $G$, all the initially estimated mask values were close to 0.5). Figure~\ref{fig:demo_different_score} shows an example of assigning different $s$ to $G$, and the learning process evaluated on the validation set is also illustrated in Figure~\ref{fig:demo_different_score} (c) and (g). Compared to the generation of clean speech (the entire learning process for generating clean speech is presented in Figure~\ref{fig:learning_process}), MetricGAN can attain the desired score more easily when $s$ is small. This phenomenon is because the number of solutions decreases gradually when $s$ increases (it is easier to obtain noisy speech than a clean speech). Therefore, the solution for a large $s$ is considerably difficult to obtain. Figures~\ref{fig:demo_different_score} (d) to (f) and (h) to (j) present the generated speech by assigning different $s$ with STOI and PESQ as $Q$, respectively. Intriguingly, the speech components gradually disappear when we attempt to generate a speech with low STOI score (the speech components are almost removed as shown by the black rectangle in Figure~\ref{fig:demo_different_score} (f)). Because STOI measures the intelligibility of speech, it is reasonable that the speech component is most crucial in this metric. On the contrary, because PESQ measures the quality of speech, the generated speech with lower $s$ seems to become more noisy (for extremely low $s$ values (Figure~\ref{fig:demo_different_score} (j)), in spite of not as serious as the STOI case, there is also some speech components being removed). These results verify that the MetricGAN can generate data according to the designate metric score and make the label space of $D$ continuous.

\begin{figure*}[ht]
\vskip -0.1in
\begin{center}
\centerline{\includegraphics[width=17.5cm]{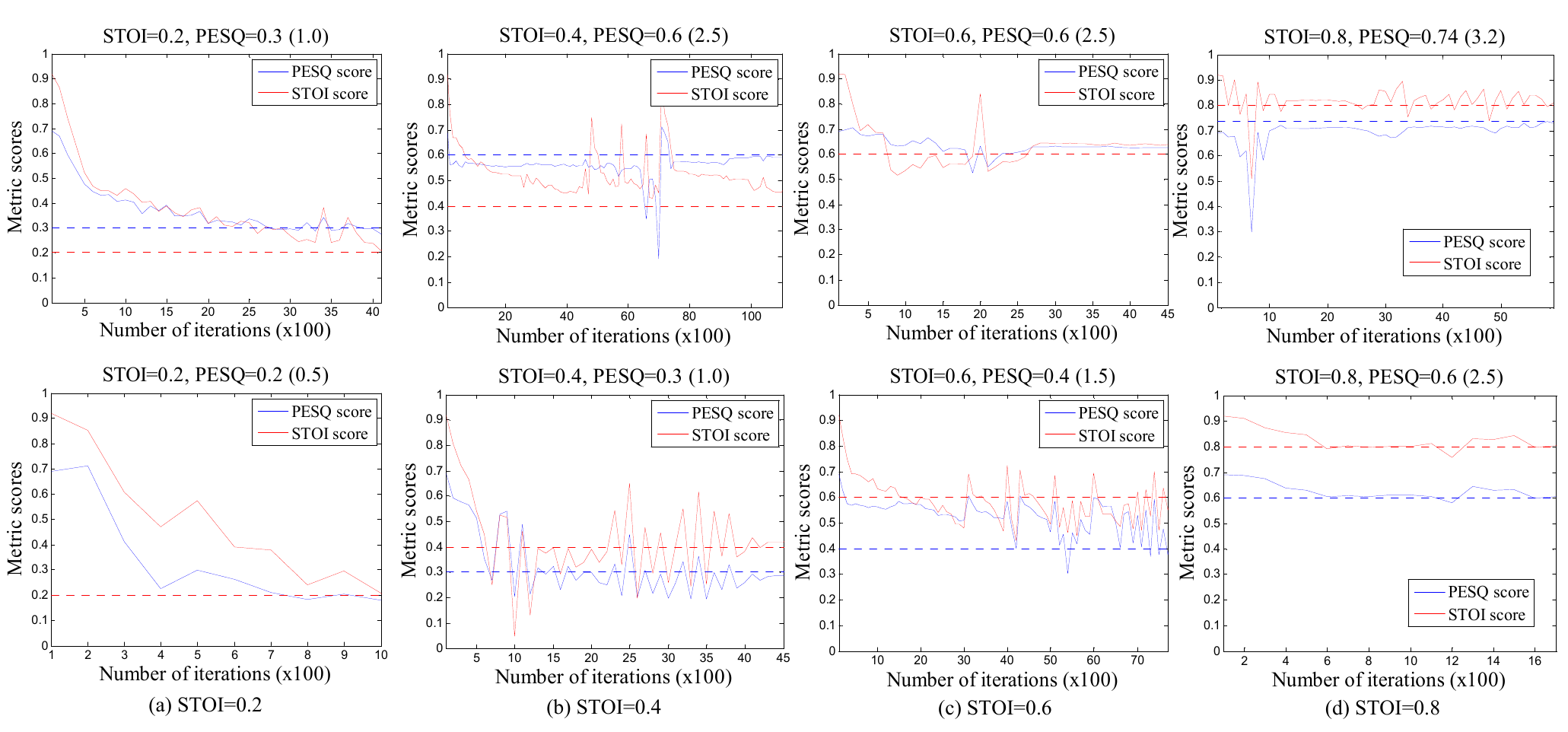}}
\vskip -0.2in
\caption{Learning curves of assigning different pairs of (STOI, PESQ) scores (shown in the title of each figure). Given a specified STOI score, the upper row and lower row is the maximum and minimum PESQ scores MetricGAN can reach, respectively. Note that the PESQ score is normalized between 0 to 1 with the original score shown in the parentheses.}
\label{fig:multi-metrics}
\end{center}
\vskip -0.2in
\end{figure*}

\begin{algorithm}[tb]
   \caption{Multi-Metric Scores Assignment}
   \label{alg:multi-metrics}
\begin{algorithmic}
   \STATE {\bfseries Input:} 
   desired score $s_1$ for metric $Q_1'(.)$ to $s_N$ for metric $Q_N'(.)$ (assume there are $N$ different metrics).
   
   \REPEAT
   \STATE 
   1) Find metric index $i$ with the largest distance between achieved and assigned score:
   
   \qquad $i=\operatorname*{argmax}_n |Q_n'(G(x),y)-s_n|$ \vspace{2mm}
   
   2) Train $G$ to minimize the loss from $D_i$:
   
   \qquad $L_{G (MetricGAN)}= \mathbb{E}_{x} [(D_i(G(x), y) - s_i)^2]$ \vspace{2mm}
   
   3) Train all $D_n$ to minimize the distance from $Q_n$: 
    
    \quad $\begin{aligned}
    L_{D (MetricGAN)}&= \mathbb{E}_{x,y} [(D_n(y, y) - 1)^2
    \\ &+(D_n(G(x), y) - Q_n'(G(x), y))^2]
    \end{aligned}$
       
       \UNTIL{converge}
    \end{algorithmic}
    \end{algorithm}

\subsubsection{Multi-Metric Scores Assignment}
In this section, we further explore the assignment of scores for multiple metrics simultaneously. Compared with single metric assignment, this is a more difficult task because the requirement to achieve other metrics can be treated as adding constraints.

Algorithm~\ref{alg:multi-metrics} shows the proposed training method for multi-metric scores assignment. Assuming that there are $N$ different metrics, we have to employee $N$ discriminators. In each iteration, only $D$ with the largest distance between achieved score, $Q_n'(G(x),y)$, and assigned score, $s_n$, would guide the learning of $G$ (steps 1 and 2). However, in the training of $D$, all the discriminators $D_n$ are updated, irrespective of whether it is used to provide loss to $G$ (step 3). 

Figure~\ref{fig:multi-metrics} shows the learning curves for the case of $N$=2. To explore more possible combinations, these results are based on the subset (top 10\% metric score) of the original validation set. To clearly illustrate the results of multi-metric learning, in each column of this figure, the assignment of STOI score is fixed with different PESQ scores. 
Because different metrics may have some positive correlation between each other, MetricGAN is difficult to converge when the score assignments are too extreme (in this case, the solution may not even exist). However, we still obtain some flexibility to generate speech with desired multiple scores. This experiment verifies that MetricGAN can approximate and distinguish different metrics well. 

\begin{table}[t]
\caption{Compared MetricGAN with other state-of-the-art methods. The highest score per metric is highlighted with bold text.}
\label{segan dataset}
\vskip 0.15in
\begin{center}
\begin{small}
\begin{tabular}{c||c||ccc}
\toprule
 & \textbf{PESQ} & CSIG & CBAK & COVL  \\
\hline
\textbf{Noisy}              & 1.97 & 3.35 & 2.44 & 2.63 \\
\textbf{SEGAN}              & 2.16 & 3.48 & 2.94 & 2.80 \\
\textbf{MMSE-GAN}          & 2.53 & 3.80 & 3.12 & 3.14 \\
\textbf{WGAN-GP}          & 2.54 & - & - & - \\
\textbf{Deep Feature Loss}                & - & 3.86 & \textbf{3.33} & 3.22 \\
\textbf{SERGAN}                & 2.62 & - & - & - \\
\hline
\textbf{MetricGAN (P)}      & \textbf{2.86} & \textbf{3.99} & 3.18 & \textbf{3.42} \\
\bottomrule
\end{tabular}
\end{small}
\end{center}
\vskip -0.1in
\end{table}




\subsection{Comparison with Other State-of-the-Art SE Models}
To further compare the proposed MetricGAN with other state-of-the-art methods, we use a publicly available dataset released by \cite{valentini2016investigating}. This dataset contains a large amount of pre-mixed noisy-clean paired data and is already used by several SE models. By using the exact same training and test dataset split, we can establish a fair comparison with them easily.

\textbf{Experimental Setup and Results:} Details about the data can be found in the original paper. Except for input features and activation functions, the network architecture and training strategy are the same as described in the previous section. In addition to the PESQ score, we also report another three metrics over the test set to compare with previous works: \textbf{CSIG} predicts the mean opinion score
(MOS) of the signal distortion, \textbf{CBAK} predicts the MOS of the background noise interferences, and \textbf{COVL} predicts the MOS of the overall speech quality, these three metrics range from 1 to 5.

Five baseline models that rely on another network to provide loss information are compared with the proposed \textbf{MetricGAN (P)}. We briefly explain these models as follows: \textbf{SEGAN} \cite{pascual2017segan} directly operates on the raw waveform and the model is trained to minimize the combination of adversarial and $L_1$ losses. \textbf{MMSE-GAN} \cite{ soni2018time} is a time-frequency masking-based method that uses a GAN objective along with $L_2$ loss. Similar to the structure of \textbf{SEGAN}, \textbf{WGAN-GP} and \textbf{SERGAN} \cite{baby2019sergan} introduced Wasserstein loss and relativistic least-square loss for GAN training, respectively.
Finally, \textbf{Deep Feature Loss}  \cite{germain2018speech} also operates on the raw waveform and is trained with a deep feature loss from another network that classifies acoustic environments. Table \ref{segan dataset} summarizes that our proposed method outperforms all previous works with respect to three metrics. This implies that although \textbf{MetricGAN} is only trained to optimize a certain score (PESQ), it also has a great generalization ability to other metrics.

\section{Conclusion}
In this paper, we proposed a novel MetricGAN approach to directly optimize generators based on one or multiple evaluation metric scores. By associating a discriminator with the metrics of interest, MetricGAN can be treated as an iterative process between surrogate loss learning and generator learning. This surrogate can successfully capture the behavior of the metrics and provides accurate gradients guiding the generator updates. In addition to outperforming other loss functions and state-of-the-art models in SE, MetricGAN can also be trained to generate data according to the designate metric scores. To the best of our knowledge, this is the first work that employs GAN to directly train the generator with respect to multiple evaluation metrics.



\nocite{langley00}

\bibliography{example_paper}
\bibliographystyle{icml2019}



\end{document}